\documentclass[prx,twocolumn,showpacs,superscriptaddress,preprintnumbers,amssymb]{revtex4-2}
\usepackage{graphicx}
\usepackage{float}
\usepackage{color}
\usepackage{dcolumn}
\usepackage{bm}
\usepackage{physics}
\usepackage{comment}
\usepackage{svg}
\usepackage{placeins}
\usepackage{hyperref}
\usepackage{soul}
\usepackage{feynmp-auto}
\usepackage{tikz}
\usepackage[normalem]{ulem} 
\usetikzlibrary{decorations.pathmorphing}
\usetikzlibrary{arrows.meta,decorations.markings}

\newcommand{\M}{{\cal{M}}}

\newcommand{\beq}{\begin{equation}}
\newcommand{\eeq}{\end{equation}}
\newcommand{\beqn}{\begin{eqnarray}}
\newcommand{\eeqn}{\end{eqnarray}}

\newcommand{\cA}{ {\cal A} }
\newcommand{\cB}{ {\cal B} }

\newcommand{\cG}{ {\cal G} }

\newcommand{\cK}{ {\cal K} }
\newcommand{\cL}{ {\cal L} }

\newcommand{\cZ}{ {\cal Z} }

\newcommand{\sfigref}[2]{Fig.~\hyperref[#1]{\ref{#1}#2}}

\begin{document}

\title{Strong-to-Weak Spontaneous Symmetry Breaking of Dephased Fermions}

\author{Abhijat Sarma}

\affiliation{Department of Physics, University of California,
Santa Barbara, CA 93106, USA}

\begin{abstract}
Strong-to-weak spontaneous symmetry breaking (SWSSB) is a novel phase transition in mixed states associated with the loss of global charge information. In this work we study fermions with $U(1)$ symmetry under infinite density dephasing and derive the SWSSB properties of the resulting mixed states. Our central observation is that at infinite dephasing every Renyi correlator is exactly the spin-spin correlator of a compact XY model, whose Boltzmann weight is the squared full-counting-statistics generating function of a classical ensemble built from the parent state. For the Renyi-2 correlator it is the full counting statistics of the parent state itself. We complement the infinite dephasing mapping with a novel diagrammatic expansion for the Renyi-1 correlator, applicable to any diagonal mixed state, which computes the couplings of the Renyi-1 XY model. We also study a replica field theory for even Renyi index quantities, applicable at finite dephasing. Our analysis indicates that fully dephased metals display long-range SWSSB in $d\geq 2$, as confirmed numerically and experimentally in $d=2$, and quasi-long-range SWSSB in $d=1$. For insulating parent states both an SWSSB and a trivial phase are possible at infinite dephasing depending on UV details. Long-range SWSSB in $d\geq 3$, or quasi-long-range SWSSB in $d=2$, survives above a nonuniversal, Renyi-index-dependent threshold and is lost when the parent state is deep in the insulating phase, with the exception of quantum Hall insulators which generically have quasi-long-range SWSSB. For even Renyi indices we also determine the stability of the SWSSB phase away from infinite dephasing and derive the universality class of the SWSSB transition.
\end{abstract}

\maketitle

\tableofcontents

\section{Introduction}

Spontaneous symmetry breaking (SSB) is a unifying theme in condensed matter physics, giving a systematic classification of many equilibrium phases of matter and transitions between them, known as the Landau paradigm \cite{Landau_1937}. Though many exotic beyond-Landau phases have been identified in the equilibrium setting including topological orders \cite{Wen_1990} and deconfined criticality \cite{Senthil_2004}, an equally intriguing direction is generalizing our understanding of phases of matter to out-of-equilibrium, open quantum systems which do not have a Hamiltonian description. It has become clear that in these open settings, density matrices can display so-called \textit{information theoretic} phase transitions, invisible to local correlation functions but detectable by quantities nonlinear in the density matrix such as entanglement entropy. Such mixed-state phases have received considerable attention recently, including measurement-induced phase transitions \cite{Li_2018, Skinner_2019, Li_2019, Chan_2019, Gullans_2020, Choi_2020, Bao_2020, Jian_2020, Potter_2022_Review, Fisher_2023_Review}, decodability transitions \cite{Dennis_2002, Fan_2024, Bao_2026}, learnability transitions \cite{Agrawal_2022, Barratt_2022_Sharpening, Barratt_2022_Learnability, Dehghani_2023, Ippoliti_2024}, information critical phases \cite{Lee_2023, Zou_2023, Garratt_2023, Weinstein_2023, Murciano_2023, Yang_2023, Kaixiang_CFT, Ashida_2024, Sang_2024, Abhi_CFTChern}, etc.

An important step in the investigation of mixed-state phases of matter is understanding how symmetries and SSB are realized in the mixed state setting. It turns out that mixed-states can possess two inequivalent forms of symmetry known respectively as \textit{strong} and \textit{weak} symmetry \cite{Buca_2012, Albert_2014, de_Groot_2022, Lee_2025, Ma_Wang_2023, Ruochen_2025}, characterized by whether or not the symmetry charge sector fluctuates between different states in the ensemble or not. Accordingly, a new form of symmetry breaking known as strong-to-weak spontaneous symmetry breaking (SWSSB) is possible \cite{Lee_2023_Hydro, Leo_SWSSB, Sala_2024, Gu_2025, Huang_2025, Weinstein_Renyi1, Zhang_2025_OneForm, Guo_Yang_2025, Kuno_2025, Kim_2026_SYK, Jake_Hydro, Lee_2026, Leo_LocalSWSSB, Chong_SWSSB}, associated with a loss of ability to reconstruct the global charge information from the local density matrix on a subregion. A particularly natural setting for investigating such physics is decohered quantum many-body ground states \cite{Lee_2025, Lee_2023, Fan_2024, Kaixiang_CFT, Bao_2026, Kaixiang_SpinLiquid, Abhi_CFTChern, FQHdecoherence, Abhi_Inequality, Si_SWSSBObs, Guo_2025_CriticalSWSSB, Lu_2026}. In other words, one seeks to understand the fate of a quantum phase of matter after letting it interact with an environment without exchanging charge with the environment. The resulting mixed state describing the system after tracing out the environment often displays SWSSB, though a generic theory akin to the Landau-Ginzburg theory for pure state SSB has not been developed.

Much of the existing work on decohered ground states has focused on ground states with severe kinematic constraints like anomaly \cite{Moharramipour_2024, Lessa_2025_Anomaly, Wang_2025_IntrinsicMSTO, Xu_Jian_2025} or particularly tractable field theoretic descriptions such as topological orders \cite{Fan_2024, Chen_Grover_2024a, Chen_Grover_2024b, Hwang_2024, Sohal_2025, Ellison_2025, Zhang_2025_OneForm, Bao_2026}, symmetry-protected topological phases \cite{de_Groot_2022, Lee_2025, Ma_Wang_2023, Ruochen_2025, Ma_Turzillo_2025, Guo_Yang_2025, Lu_2026}, and conformal field theories \cite{Lee_2023, Zou_2023, Garratt_2023, Weinstein_2023, Murciano_2023, Yang_2023, Kaixiang_CFT, Ashida_2024, Abhi_CFTChern, Guo_2025_CriticalSWSSB}, often employing techniques such as fixed-point wavefunctions \cite{Fan_2024, Chen_Grover_2024a, Sohal_2025, Ellison_2025, Bao_2026, FQHdecoherence} and boundary conformal field theory \cite{Lee_2023, Zou_2023, Garratt_2023, Kaixiang_CFT, Abhi_CFTChern} as convenient computational tools. On the other hand, this work focuses on a wide class of fermionic systems encompassing both the metallic (Fermi liquid) and insulating phases, using generic methods that do not require a fine-tuned ground state and are applicable even beyond the fermionic setting considered in this work.

Our technical contributions are threefold. First, we show that at infinite dephasing every Renyi correlator is exactly the spin-spin correlator of a compact XY model. Its Boltzmann weight is the squared full-counting-statistics generating function of a classical ensemble $p^{\alpha/2}$ built from the Born probabilities $p$ of the parent state, with $\alpha$ being the Renyi index. For the Renyi-2 correlator this is the full counting statistics of the parent state itself. This representation yields a rigorous lower bound on the Renyi-1 and fidelity correlators in terms of the correlation function of the parent state, and allows us to qualitatively understand the phase diagram at infinite dephasing. Second, we show that finite dephasing acts on the XY model as an explicit symmetry-breaking field, and we derive a replica field theory of the doubled state to study the finite-dephasing problem. This replica theory requires a subtle regularization, which is investigated in more detail in another work \cite{Abhi_Gutz}. Third, we introduce a novel expansion for Renyi-1 correlation functions of diagonal density matrices, i.e. classical distributions. This allows us to explicitly compute the couplings in the Renyi-1 XY model. This expansion can also be applied outside of the setting of decohered quantum ground states and be utilized to study any classical data.

Our main physical conclusions are as follows. For Luttinger liquids and Laughlin states we can compute the Renyi correlators at infinite dephasing exactly, which are algebraically decaying. In general, the renormalized kernel of the XY model has the same IR form as the parent state's equal-time structure factor $S(q)$. This allows us to conclude that maximally dephased metals display long-range SWSSB in $d\geq2$, as has been confirmed numerically \cite{Kaixiang_SpinLiquid} and experimentally \cite{Si_SWSSBObs} in $d=2$. On the other hand, for maximally dephased insulators both a SWSSB and a trivial phase are possible depending on UV details. The SWSSB phase is long-range in $d\geq 3$ and algebraic (quasi-long-range) in $d=2$. We expect the SWSSB phase to be realized close to a metallic transition where the correlation length is large, while deep in the insulating regime the strong symmetry is preserved. For insulators in $d=2$ we derive a necessary bound on the gap in terms of the kinetic energy for quasi-long-range Renyi-1 SWSSB. (Fractional) quantum Hall insulators are exempt from the threshold, and generically display quasi-long-range SWSSB \cite{Abhi_CFTChern, FQHdecoherence}. Finally, we show for even Renyi indices that long-range SWSSB persists at large but finite dephasing, while quasi-long-range SWSSB persists at large but finite dephasing only if the power-law exponent $\eta>4$. We also derive the universal Ginzburg-Landau theory of the transition into the SWSSB phase.

The rest of the work is structured as follows. In Section~\ref{sec:diagnostics} we introduce the strong and weak symmetries, strong-to-weak spontaneous symmetry breaking, and the order parameters which diagnose it, i.e.\ the Renyi correlators. In Section~\ref{sec:dephase} we define the setup of dephased fermionic ground states which are the object of study in the rest of the work. In Section~\ref{sec:xy} we derive the exact XY representation of the Renyi correlators at infinite dephasing. In Section~\ref{sec:bosonization} we compute the Renyi correlators for Gaussian distributions. In Section~\ref{sec:phases} we analyze the phases of the XY model at infinite dephasing for metallic, insulating and quantum Hall parent states. In Section~\ref{sec:replica} we turn to finite dephasing and construct a replica field theory. In Section~\ref{sec:diagrams} we introduce a diagrammatic expansion of Renyi-1 correlators of classical distributions in terms of the connected correlation functions of the distribution, which computes the couplings of the Renyi-1 XY model. We conclude in Section~\ref{sec:conclusions}, summarizing our results and looking towards future directions.

\section{Diagnostics of SWSSB}
\label{sec:diagnostics}

In the mixed state setting, the conventional notion of symmetry is enriched into two distinct classes, namely the \textit{strong} and \textit{weak} symmetries \cite{Buca_2012, Albert_2014, de_Groot_2022, Ma_Wang_2023}. In the case of a $U(1)$ symmetry with symmetry operator $U_{\theta}$, a density matrix $\rho$ is strongly symmetric iff
\begin{equation}
    U_{\theta} \rho = e^{i\theta N}\rho
\end{equation}
The interpretation of the strong symmetry is that $\rho$ is an ensemble of states that each possess $U(1)$ charge $N$. On the other hand, the weak symmetry is characterized by the weaker condition
\begin{equation}
    U_{\theta} \rho U_{\theta}^\dagger = \rho
\end{equation}
which corresponds to $\rho$ being an ensemble of $U(1)$ symmetric states  with different charges. One can consider a strongly symmetric $\rho$ to be akin to a canonical ensemble of $U(1)$ charge, while a weakly symmetric $\rho$ is akin to a grand canonical ensemble, though of course $\rho$ need not be a Gibbs state in general. 

Just as symmetries can be spontaneously broken in the pure state case, so too can the strong and weak symmetries be broken spontaneously. The conventional notion of symmetry breaking, which we refer to as strong-to-nothing and weak-to-nothing spontaneous symmetry breaking, are diagnosed by ordinary correlation functions, by which we mean correlation functions linear in the density matrix, i.e.
\begin{equation}
    C(x, y) = \tr[\rho c^\dagger_x c_y]
\end{equation}
where $c_x$ is an order parameter charged under the $U(1)$ symmetry, here taken to be a fermionic annihilation operator. Saturation of this correlation function to a nonzero constant at large distances implies that the $U(1)$ charge is long-range ordered, which is a stable property of a phase of matter in accordance with the Landau paradigm. 

Similarly, the strong symmetry can be spontaneously broken to weak, dubbed strong-to-weak spontaneous symmetry breaking (SWSSB) \cite{Lee_2023_Hydro, Leo_SWSSB, Sala_2024, Gu_2025, Huang_2025}. For example, the equivalence of the canonical and grand-canonical Gibbs states in the thermodynamic limit can be viewed as SWSSB of the canonical Gibbs state \cite{Leo_SWSSB, Sala_2024}. One interpretation is that in the SWSSB phase, one cannot distinguish the mixed state $\rho$ from another mixed state $\rho' = c^\dagger_x c_y \rho c_y^\dagger c_x$ with a charge displaced from $y$ to $x$. Thus SWSSB should be viewed as an information theoretic property of $\rho$ \cite{Leo_SWSSB, Chong_SWSSB}. Since the conventional correlation function is sensitive to breaking of both the strong and weak symmetries, detecting the SWSSB necessitates an unconventional kind of order parameter that is nonlinear in the density matrix. Indeed the original proposal for detecting SWSSB was via the \textit{fidelity} correlator \cite{Leo_SWSSB},
\begin{equation}
    C_{F}(x, y) = F(\rho, c^\dagger_x c_y \rho c_y^\dagger c_x) = \tr\sqrt{\sqrt{\rho} c^\dagger_x c_y \rho c_y^\dagger c_x \sqrt{\rho}}
\end{equation}
with the fidelity $F(\rho, \rho')$ defining a natural notion of distinguishability between $\rho$ and $\rho'$ \cite{Nielsen_Chuang}.

Like the conventional correlation function, long range order of the fidelity correlator is a stable property of mixed-state phases in the sense of being preserved by finite-depth locally reversible channels \cite{Coser_2019, Leo_SWSSB, Sang_Hsieh_2025, Shengqi_Phases}, and thus serves as a sensible definition of SWSSB as a property of a mixed-state phase. However, it is often inconvenient for analytic treatment. Instead, one often considers the \textit{Renyi-$\alpha$ two-point correlators},
\begin{equation}
    C^{(\alpha)}(x, y) = \frac{\tr[\rho^{\alpha/2} c^\dagger_x c_y \rho^{\alpha/2} c^\dagger_y c_x]}{\tr[\rho^\alpha]}
\end{equation}
which are more amenable to analytic treatment. Indeed the Renyi-1 correlator enjoys a similar stability theorem as the fidelity correlator, and is thus an equally good measure of SWSSB, thanks to the inequality \cite{Weinstein_Renyi1}
\begin{equation}
\label{eqn:fidelity_ineq}
    C_{F}(x, y)^2 \leq C^{(1)}(x, y) \leq C_{F}(x, y)
\end{equation}

The Renyi-$\alpha$ correlators can be viewed as suitable replica generalizations of the Renyi-$1$ correlator, which are often even easier to calculate analytically. Indeed the Renyi-$1$ correlator can be viewed as a conventional correlation function of the \textit{canonical purification} $|\sqrt{\rho}\rangle\rangle$ of $\rho$, while the Renyi-$2$ correlator can be viewed as a conventional correlation function of the \textit{Choi double} $|\rho\rangle\rangle$ of $\rho$. Even more generally, we can consider the \textit{asymmetric} Renyi two-point correlators
\begin{equation}
\label{eqn:asym_renyi}
    C^{(k, k')}(x, y) = \frac{\tr[\rho^{k/2}c^\dagger_x c_y \rho^{k'/2} c^\dagger_y c_x]}{\tr[\rho^{{(k+k')/2}}]}
\end{equation}
We refer to $\alpha \in \mathbb{Z}$ and $(k+k')/2 \in \mathbb{Z}$ interchangeably as the Renyi index; the letter $n$ is reserved for occupation configurations and densities. We will introduce various techniques for computing these quantities that are applicable for different values of the Renyi index, though we emphasize that the Renyi-1 correlator is the true diagnosis of SWSSB, with the other Renyi quantities only serving as proximal probes that do not come with any stability theorem \cite{Leo_SWSSB}.

It was recently proposed that one can define a \textit{local} analogue of SWSSB \cite{Leo_LocalSWSSB} using the reduced density matrix $\rho_A = \tr_{\bar{A}}[\rho]$ on a region $A$ containing a point $x$. Then, the \textit{Renyi-$\alpha$ one-point correlator}
\begin{equation}
\label{eqn:renyi_1pt}
    C^{(\alpha)}(x, A) = \frac{\tr[\rho_A^{\alpha/2} c^\dagger_x \rho_A^{\alpha/2} c_x]}{\tr[\rho_A^{\alpha}]}
\end{equation}
probes SWSSB of $\rho$ on the region $A$, which has the interpretation that $\rho$ and $\rho'=c^\dagger_x \rho c_x$ cannot be distinguished using data on $A$ alone. Accordingly, global SWSSB as diagnosed by the Renyi-1 two-point function necessarily implies local SWSSB, but the converse is not true. 

Lastly, another useful diagnostic of mixed state phases is the conditional mutual information (CMI) \cite{Sang_2024, Sang_Hsieh_2025, Lessa_2025_Anomaly, Chong_SWSSB}, defined as a difference of two mutual informations (MI)
\begin{align}
\begin{split}
    I(A:C|B) &= I(A,BC) - I(A,B) \\&= S(AB)+S(BC)-S(B)-S(ABC)
\end{split}
\end{align}
where $S(A) = -\tr[\rho_A \log(\rho_A)]$ is the Von-Neumann entropy. The CMI captures the failure of $\rho$ to be a \textit{quantum Markov chain} \cite{Hayden_2004}. In other words, if the CMI is small, then $\rho$ can approximately be recovered from $\rho_{AB}$ via a recovery channel acting on region $B$ alone. This is quantified by the recoverability relation \cite{Fawzi_2015}
\begin{equation}
    I(A:C|B) \geq -2\log F(\rho, \mathcal{R}_{B\rightarrow BC} [\rho_{AB}])
\end{equation}
where $\mathcal{R}_{B\rightarrow BC}$ is some optimal recovery map.

Both global and local SWSSB imply long-range CMI \cite{Leo_LocalSWSSB, Lee_2026, Chong_SWSSB}. This can be intuitively understood as meaning that when SWSSB occurs, the charge information on $AB$ does not contain enough information to recover the charge of the entire system, as we cannot distinguish $\rho_A$ from $\rho'_A$ where $\rho'$ has a charge displaced from $y \in \bar A$ to $x \in A$.

\section{Dephased Fermions}
\label{sec:dephase}
In this work, we consider the SWSSB properties of dephased fermions with $U(1)$ symmetry. We consider a fermionic many-body ground state $|\psi_0\rangle$, with density matrix $\rho_0 = |\psi_0\rangle \langle \psi_0|$. We denote $N$ the total fermion number of $|\psi_0\rangle$, i.e. the $U(1)$ charge of $|\psi_0\rangle$, and $\Lambda$ the set of lattice sites, with $|\Lambda| = N_s$. We are interested in the long-time limit of strongly symmetric Lindbladian dynamics \cite{Lindblad_1976, GKS_1976} on this starting state, i.e.

\begin{equation}
\rho = \lim_{t\rightarrow \infty}e^{t\mathcal{L}} \rho_0
\end{equation}
The Lindbladian dynamics generated by $\mathcal{L}$ are specified by some choice of $U(1)$ symmetric ``jump" operators $\{L_i\}$. There are a few choices one might make. One choice is the hopping operators $L_{ij} = c^\dagger_i c_j$. This leads to very simple dynamics in the long-time limit, however; the jump operators are noncommuting, and the only steady state is the maximally mixed state (in the fixed charge sector set by $|\psi_0\rangle$), which trivially exhibits SWSSB \cite{Leo_SWSSB, Jake_Hydro}. The other obvious choice is the onsite densities $L_i = n_i$, which are commuting and can lead to nontrivial steady-states \cite{Kaixiang_SpinLiquid, Abhi_CFTChern, FQHdecoherence, Abhi_Inequality, Si_SWSSBObs}. Understanding the SWSSB properties of these steady-states is the focus of this work. The infinite dephasing limit is easier to analyze because the density matrix becomes purely classical (diagonal),

\begin{equation}
\label{eqn:diag_ensemble}
    \rho = \sum_{n} p\left[n\right] |n\rangle \langle n|
\end{equation}
where $|n\rangle = \prod_{z|n_z = 1} c^\dagger_{z} |0\rangle$ denotes the product state with fermion occupation given by $n = \{n_z\} \in \mathbb{Z}_2^{N_s}$ with $\sum_z n_z = N$. The configuration probability is obtained from the Born rule as

\begin{equation}
\label{eqn:config_probs}
p[n] = |\langle n | \psi_0\rangle|^2
\end{equation}
Indeed the effect of the dephasing can be understood as suppressing the off-diagonal elements of $\rho$ in the occupation basis while leaving the diagonal elements unchanged. Because of this, all density correlation functions of $\rho$ are inherited from $|\psi_0\rangle$, which allows us to discern the universal properties of $\rho$ from those of $|\psi_0\rangle$.

Diagonality has a second consequence that will be important throughout. Since $\rho$ commutes with every $n_z$, it is invariant under \textit{site-dependent} $U(1)$ rotations,
\begin{equation}
\label{eqn:weak_local}
    e^{i\sum_z\theta_z n_z}\,\rho\,e^{-i\sum_z\theta_z n_z}=\rho\qquad\text{for all }\{\theta_z\}
\end{equation}
a \textit{weak local} $U(1)$ symmetry \cite{Abhi_Gutz, Kaixiang_SpinLiquid, Jake_Hydro}, of which the weak global symmetry is the uniform case. It is exact only at infinite dephasing. This is what allows the exact representation of Sec.~\ref{sec:xy}, and it is what confines the vortices of quantum Hall insulators in Secs.~\ref{sec:hall} and \ref{sec:hall_replica}.

For later use, we define the connected $k$-point density correlation functions
\begin{equation}
\label{eqn:Sc_def}
    S_c^{(k)}(z_1, \dots, z_k) = \langle \psi_0 | n_{z_1} \dots n_{z_k} | \psi_0 \rangle_c
\end{equation}
and their Fourier transforms $S_c^{(k)}(q_1, \dots q_k)$. We identify the filling $\nu = S_c^{(1)}(z)$ as well as the equal-time structure factor $S(q)$ by the relation
\begin{equation}
\label{Sq_def}
    \langle \psi_0 | n_{z_1} n_{z_2}| \psi_0 \rangle_c = \int \frac{d^d q}{(2\pi)^d}S(q) e^{iq\cdot (z_1-z_2)}
\end{equation}
or equivalently
\begin{equation}
S_c^{(2)}(q_1, q_2) = S(q_1) (2\pi)^d \delta^{d}(q_1+q_2)
\end{equation}
Momentum integrals are understood to be over the Brillouin zone throughout. 

Lastly, we note that the $U(1)$ symmetry implies that $\langle \psi_0| \int d^dz \, \delta n(z) \cdot X | \psi_0\rangle = 0$ where $\delta n(z) = n(z) - \nu$ for any functional $X=X[n]$ of the density configuration, which forces $S_c^{(k)}$ to vanish when any of its arguments go to zero,
\begin{equation}
\label{eqn:Sc_charge_neutral}
S^{(k)}_c(q_1, \dots, q_i \rightarrow 0, \dots, q_k) = 0
\end{equation}

\section{Exact XY Representation at Infinite Dephasing}
\label{sec:xy}

At infinite dephasing the density matrix is the classical ensemble of Eq.~\ref{eqn:diag_ensemble}, i.e. it has the weak local symmetry of Eq.~\ref{eqn:weak_local}, and every Renyi quantity is a sum over occupation configurations. In this section we show that these sums are exactly the correlators of a compact XY model whose Boltzmann weight is built from the full counting statistics of a classical ensemble derived from the parent state. The representation is exact on the lattice and applies to any diagonal density matrix. Throughout, $\alpha$ denotes the Renyi index and $n=\{n_z\}$ an occupation configuration.

\subsection{Boson representation}

Define $\cZ_\beta=\sum_n p[n]^\beta=\tr\rho^\beta$ and the family of normalized hard-core boson states
\begin{equation}
\label{eqn:boson_state}
|\Phi_\alpha\rangle = \cZ_\alpha^{-1/2}\sum_n p[n]^{\alpha/2}\,|n\rangle_b
\end{equation}
where $|n\rangle_b$ is the bosonic Fock state with the same occupations as $|n\rangle$. Since $|\langle n'|c^\dagger_x c_y|n\rangle|^2\in\{0,1\}$, the fermionic sign drops out of the Renyi correlators of a diagonal ensemble, and one finds
\begin{align}
\begin{split}
\label{eqn:boson_odlro}
C^{(\alpha)}(x,y) &= \langle\Phi_\alpha|b^\dagger_x b_y|\Phi_\alpha\rangle \\
C^{(k,k')}(x,y) &= \frac{\sqrt{\cZ_k\cZ_{k'}}}{\cZ_{(k+k')/2}}\,\langle\Phi_k|b^\dagger_x b_y|\Phi_{k'}\rangle
\end{split}
\end{align}
In words, SWSSB of the dephased state as diagnosed by the Renyi-$\alpha$ correlator is off-diagonal long-range order of the positive boson wavefunction $|\psi_0[n]|^{\alpha}$. The same holds for the one-point function of Eq.~\ref{eqn:renyi_1pt}. With $p_A$ the marginal distribution on $A$,
\begin{align}
\begin{split}
\label{eqn:boson_1pt}
C^{(\alpha)}(x,A) &= \langle\Phi_{\alpha,A}|b^\dagger_x|\Phi_{\alpha,A}\rangle \\
|\Phi_{\alpha,A}\rangle&\propto\sum_{n_A}p_A[n_A]^{\alpha/2}|n_A\rangle_b
\end{split}
\end{align}
so that local SWSSB is literally the condensate amplitude of the marginal boson state, which is a superposition of different particle numbers. Different Renyi indices probe different members of the family $|\Phi_\alpha\rangle$, and long-range order of one member does not imply it for another. For $\alpha=1$ the boson wavefunction is $|\psi_0[n]|$, the parent wavefunction with its signs removed. 

Eq.~\ref{eqn:boson_odlro} also yields a rigorous bound. Let $h=\delta_x-\delta_y$ denote the transfer of one fermion from $y$ to $x$, so that $n+h$ is the configuration obtained from $n$ by moving a fermion from $y$ to $x$; this notation is used throughout. By the triangle inequality,
\begin{equation}
\label{eqn:C1_bound}
C^{(1)}(x,y) = \sum_n |\psi_0[n]|\,|\psi_0[n+h]| \geq \left|\langle\psi_0|c^\dagger_x c_y|\psi_0\rangle\right|
\end{equation}
since the fermionic signs of $\langle n+h|c^\dagger_x c_y|n\rangle$ can only reduce the right-hand side. Together with $C_F\ge C^{(1)}$, Eq.~\ref{eqn:C1_bound} states that neither the Renyi-1 nor the fidelity correlator of the dephased state can decay faster than the correlation function of the parent state. For a Fermi liquid the latter is a power law, $|\langle c^\dagger_y c_x\rangle|\sim r^{-(d+1)/2}$, so a dephased metal is never short-range correlated in the Renyi-1 or fidelity sense, in any dimension: the only question is whether the SWSSB is long-range or quasi-long-range. No such bound exists for $\alpha\geq 2$.

\subsection{Phase representation, spin waves, and vortices}

We now pass to variables conjugate to the occupations. Let $\varrho_\alpha[n]=p[n]^{\alpha/2}/\cZ_{\alpha/2}$ be the normalized ``Renyi ensemble'' and
\begin{equation}
\label{eqn:Psi_def}
\Psi_\alpha[\theta] = \sum_n \varrho_\alpha[n]\, e^{i\theta\cdot n} = \left\langle e^{i\theta\cdot n}\right\rangle_{\varrho_\alpha}
\end{equation}
its full counting statistics generating function, with $\theta\cdot n=\sum_z\theta_z n_z$ and $\theta_z\in[0,2\pi)$. For $\alpha=2$ the Renyi ensemble is the parent distribution itself, and $\Psi_2[\theta]=\langle\psi_0|e^{i\theta\cdot n}|\psi_0\rangle$ is the full counting statistics generating function of the parent state; for $\alpha=1$ it is the full counting statistics of $\sqrt p$. Inserting $\delta_{n,n'}=\int D\theta\,e^{i\theta\cdot(n-n')}$, with $D\theta=\prod_z d\theta_z/2\pi$, into Eq.~\ref{eqn:boson_odlro} gives
\begin{align}
\begin{split}
\label{eqn:xy_corr}
C^{(\alpha)}(x,y) &= \frac{\int D\theta\, e^{-S_\alpha[\theta]}\, e^{i(\theta_x-\theta_y)}}{\int D\theta\, e^{-S_\alpha[\theta]}} \\
e^{-S_\alpha[\theta]}&\equiv|\Psi_\alpha[\theta]|^2
\end{split}
\end{align}
i.e. the Renyi-$\alpha$ correlator is the spin-spin correlator of a classical model of planar spins $e^{i\theta_z}$ with Boltzmann weight $|\Psi_\alpha|^2$. Writing $W_\alpha[J]=\log\sum_n\varrho_\alpha[n]e^{J\cdot n}$ for the cumulant generating function of the Renyi ensemble,
\begin{equation}
\label{eqn:S_alpha}
S_\alpha[\theta] = -2\,\mathrm{Re}\, W_\alpha[i\theta]
\end{equation}
Three properties of $S_\alpha$ are exact. (i) It is $2\pi$-periodic in each $\theta_z$ separately, because $n_z$ is an integer. (ii) It is invariant under a global shift $\theta_z\to\theta_z+\gamma$, because $W_\alpha[J+\gamma]=W_\alpha[J]+\gamma N$ at fixed particle number. (iii) It is nonnegative, $S_\alpha\geq 0$, with the minimum attained on uniform configurations, because $|\langle e^{i\theta\cdot n}\rangle|\le 1$. The model is therefore a compact $U(1)$ (XY) model with a ferromagnetic ground state, and Eq.~\ref{eqn:xy_corr} states that SWSSB at infinite dephasing is ferromagnetic order of this XY model, long-range or quasi-long-range.

Expanding $W_\alpha$ in the cumulants $\kappa^{(\alpha)}_k$ of the density in the Renyi ensemble, only the even cumulants survive the real part in Eq.~\ref{eqn:S_alpha}:
\begin{align}
\begin{split}
\label{eqn:S_alpha_cumulant}
S_\alpha[\theta] &= \sum_{z,z'}\kappa^{(\alpha)}_2(z,z')\,\theta_z\theta_{z'} \\&\quad- \frac{1}{12}\sum \kappa^{(\alpha)}_4\,\theta^4 + \frac{1}{360}\sum\kappa^{(\alpha)}_6\,\theta^6 - \dots
\end{split}
\end{align}
where $\sum\kappa_k\theta^k$ is shorthand for $\sum_{z_1,\dots,z_k}\kappa_k(z_1,\dots,z_k)\theta_{z_1}\cdots\theta_{z_k}$. For $\alpha=2$ the $\kappa^{(2)}_k=S^{(k)}_c$ are the connected density correlators of the parent state, Eq.~\ref{eqn:Sc_def}. For $\alpha=1$ they are the cumulants of the ensemble $\sqrt p=e^{-\Gamma/2}$ with $\Gamma=-\log p$. These cumulants are computed by the expansion of Sec.~\ref{sec:diagrams}. In momentum space the quadratic term reads $\int_q \kappa_2^{(\alpha)}(q)|\theta_q|^2$, with $\kappa_2^{(2)}(q)=S(q)$ exactly. For $\alpha=1$ $\kappa_2^{(1)}(q)$ inherits the same small-$q$ form as $S(q)$ as well up to numerical prefactors, as follows from the power counting of Secs.~\ref{sec:power_counting} and \ref{sec:diagrams}.

The expansion Eq.~\ref{eqn:S_alpha_cumulant} of $S_{\alpha}$ in powers of $\theta$, which we refer to as the spin-wave expansion, is blind order-by-order to the compactness of $\theta$. The expansion therefore misses the vortex sector of the theory, which is only reconstructed at infinite order. We will use the expansion to argue for the long-wavelength form of the kinetic term of the XY model, which we will use to determine when a long-range ordered phase is destabilized by Goldstone fluctuations. Whether vortices proliferate and disorder the system is a nonperturbative question about the compact model that the expansion in $\theta$ cannot answer at finite order. We return to this point in Secs.~\ref{sec:phases} and \ref{sec:diagrams}.

For free fermions we can write $S_{2}[\theta]$ exactly. With $G_{zz'}=\langle\psi_0|c^\dagger_{z'}c_z|\psi_0\rangle$ the correlation matrix, the full counting statistics is a determinant \cite{Levitov_1996, Klich_2003}, $\Psi_2[\theta]=\det[1+(e^{i\theta}-1)G]$, so that
\begin{equation}
\label{eqn:S2_free}
S_2[\theta] = -2\,\mathrm{Re}\,\tr\log\left[1+(e^{i\theta}-1)G\right]
\end{equation}
For a product state $G$ is diagonal with $\nu_z\in\{0,1\}$, the argument of the logarithm is a pure phase, and $S_2\equiv 0$. The weight is flat and the XY model is completely disordered, as it must be, since an atomic insulator carries no information about where a transferred charge came from. 

\subsection{Charge representation}
\label{sec:spinwave}

Inserting the definition of $\Psi_\alpha$ into $|\Psi_\alpha|^2$ gives the Fourier series
\begin{align}
\begin{split}
\label{eqn:charge_rep}
e^{-S_\alpha[\theta]} &= \sum_{h\,:\,\sum_z h_z=0} \cB_\alpha[h]\, e^{-i\theta\cdot h}\\ \cB_\alpha[h]&=\sum_n\varrho_\alpha[n]\,\varrho_\alpha[n+h]\geq 0
\end{split}
\end{align}
over integer-valued charge-transfer configurations $h$. The Fourier coefficients are the multi-charge-transfer Renyi correlators of the dephased state. For $h=\delta_{x}-\delta_{y}$ a single dipole, $\cB_\alpha[h]\propto C^{(\alpha)}(x,y)$ itself. This is the analogue of the charge representation of the XY model \cite{Villain_1975, Jose_1977}. In the charge representation the spin-wave expansion becomes the expansion of $-\log \cB_{\alpha}[h]$ in powers of $h$, which is the organizing principle of Sec.~\ref{sec:diagrams}.

\section{Gaussian Distributions}
\label{sec:bosonization}
If the density distribution at long wavelengths takes the simple gaussian form
\begin{equation}
\label{eq:prob_gauss}
    p[n] = \exp(-\frac{1}{2} \int \frac{d^dq}{(2\pi)^d} \frac{|n_q|^2}{S(q)})
\end{equation}
then we can compute the Renyi correlators and entropy / CMI exactly. This is equivalent to the statement that $S_c^{(k)}$ are subleading in the IR compared to $S_c^{(2)}$ for $k>2$. This is true for Luttinger liquids as seen from bosonization as explained in Sec.~\ref{sec:luttinger}. For Laughlin states this follows from the Coulomb gas mapping and the irrelevance of vortices for quantum Hall states as explained in Sec.~\ref{sec:hall}. Note that the density distribution being gaussian is not equivalent to the parent state being noninteracting. For example, free fermions in $d\geq 2$ violate this condition.

The Renyi two-point function is
\begin{align}
\begin{split}
    C^{(k, k')}(x, y) &= \frac{\sum_n \; p[n]^{k/2} p[n+h]^{k'/2}}{\sum_n\; p[n]^{(k+k')/2}}
\end{split}
\end{align}
where $h = \delta_x - \delta_y$, with Fourier transform $h_q = e^{-iq\cdot x} - e^{-iq\cdot y}$. Then it simply follows
\begin{align}
\begin{split}
\label{eq:renyi_1d}
    &C^{(k, k')}(x, y) \\&= \frac{\sum_n\,\exp(-\frac{1}{2} \int \frac{d^dq}{(2\pi)^d}\frac{1}{S(q)} (\frac{k}{2} |n_q|^2 + \frac{k'}{2}|n_q+h_q|^2))}{\sum_n\,\exp(-\frac{1}{2} \int \frac{d^dq}{(2\pi)^d}\frac{1}{S(q)} \frac{k+k'}{2} |n_q|^2)}\\
    &= \exp(-\frac{kk'}{4(k+k')} \int \frac{d^dq}{(2\pi)^d} \frac{|h_q|^2}{S(q)})  \\
\end{split}
\end{align}

Similarly, the Renyi one-point function is easily computed. Let $A$ be a subregion containing $x$ and $\bar{A}$ its complement. Then the marginal density distribution on $A$ is
\begin{equation}
\label{eq:marginal_dist}
    p_A[n_A] = \sum_{n_{\bar{A}}} \;p[n_A + n_{\bar{A}}]
\end{equation}
Here $n_{A/\bar{A}}$ have support contained in $A/\bar{A}$ respectively. The Renyi one-point function is
\begin{align}
\begin{split}
    C^{(\alpha)}(x, A) &= \frac{\tr[\rho_A^{\alpha/2} c^\dagger_x \rho_A^{\alpha/2} c_x]}{\tr[\rho_A^{\alpha}]} \\&= \frac{\sum_{n_A} \; p_A[n_A]^{\alpha/2} p_A[n_A+\delta_x]^{\alpha/2}}{\sum_{n_A} \; p_A[n_A]^{\alpha}}\\
\end{split}
\end{align} We can rewrite the probability as a real-space Gaussian kernel
\begin{align}
\begin{split}
    p[n] &= \exp\left(-\frac{1}{2}\sum_{z,z'} \, n_{z} S^{-1}(z,z') n_{z'}\right)
\end{split}
\end{align}
with
\begin{equation}
    S^{-1}(z,z') = \int \frac{d^dq}{(2\pi)^d} \frac{e^{iq(z-z')}}{S(q)}
\end{equation}
The kernel $S^{-1}(z,z')$ is the solution to the equation
\begin{equation}
    \sum_w \, S(z,w)S^{-1}(w,z') = \delta_{z,z'}
\end{equation}
The marginal distribution on $A$ can then be written in terms of another Gaussian kernel $S_A^{-1}$ which is the inverse of the restriction $S_A$, not to be confused with the restriction of $S^{-1}$ to A. In other words,
\begin{align}
\begin{split}
    &p_A[n_A] = \exp\left(-\frac{1}{2} \sum_{z,z'\in A} n_z S_A^{-1}(z,z') n_{z'}\right)
\end{split}
\end{align}
where $S_{A}^{-1}$ is the solution to the Dirichlet problem
\begin{align}
\begin{split}
\label{eq:rest_prop}
    &\sum_{w\in A} \, S(z, w) S^{-1}_A(w,z') = \delta_{z,z'} \; \forall z,z' \in A \\
    &\qquad  S_A^{-1}(z, z') = 0 \; \; \text{if } z \text{ or } z' \notin A
\end{split}
\end{align}
In terms of the marginal kernel, the correlator is
\begin{align}
\begin{split}
\label{eq:1pt_formula_exact}
    &C^{(\alpha)}(x, A) \\
    &=\exp(-\frac{\alpha}{8} \sum_{z,z'} \, \delta_{z,x}S^{-1}_A(z,z') \delta_{z',x}) \\
    &=\exp(-\frac{\alpha}{8} S_A^{-1}(x, x))
\end{split}
\end{align}
where in the continuum expressions below the diagonal element is regulated by the lattice spacing $a$, $S_A^{-1}(x,x)\to S_A^{-1}(x,x+a)$.

Lastly, we compute the Von-Neumann entropy of $\rho_A$ as a function of $L=|A|$. The dephasing will produce a leading volume-law, but like the undephased state there will be a subleading universal logarithmic term that controls the CMI.
\begin{align}
\begin{split}
    S_{VN} &= -\sum_{n_A} \; p_A[n_A] \log p_A[n_A] \\
    &= \frac{1}{2} \log \det(2\pi e S_A)\\
\end{split}
\end{align}
The universal logarithm is then given by $\frac{1}{2}\sum_{m=1}^{N} \log \lambda_m$ where $\lambda_m$ label the eigenvalues of $S_A$ and $N \sim (L/a)^d$ is the number of modes.

\subsection{Luttinger liquid ($d=1$)}
\label{sec:luttinger}
To begin with, we address the one dimensional metallic phase, i.e. the Luttinger liquid \cite{Haldane_1981, Giamarchi_2003}. As the ground state is a CFT, the Renyi correlators before dephasing possess power-law tails. We will see that the Renyi correlators remain power law in the maximally dephased state, but with modified exponents. The structure factor is
\begin{equation}
    S(q) = \frac{K|q|}{2\pi}
\end{equation}
where $K$ is the Luttinger parameter. The noninteracting limit is given by $K=1$. The two-point correlator is
\begin{equation}
\label{eqn:ll_renyi2}
C^{(k, k')}(x, y) \sim   \left(\frac{a}{r}\right)^{\frac{kk'}{k+k'}\frac{1}{K}}
\end{equation}
where $a$ is a UV cutoff (e.g. the lattice spacing) and $r=|x-y|$. 

This result is exact for the following reason. Bosonization expresses the density as $n(x)=\nu+\partial_x\phi/\pi+\sum_{m\geq1}A_m\cos[2m(k_Fx+\phi)]$, and the Luttinger liquid ground state is Gaussian in $\phi$. The distribution of the smooth part of the density is therefore exactly of the form of Eq.~\ref{eq:prob_gauss}, with the structure factor above. The harmonics, which are what encodes the quantization of the lattice charge, give nonanalyticities of $S(q)$ and of the higher cumulants at the wavevectors $2mk_F$, which do not modify the Renyi correlator at long distances, and a compact field on a line has no vortices, so the spin-wave result is exact. Eq.~\ref{eqn:ll_renyi2} at $k=k'=\alpha$ identifies $|\Phi_\alpha\rangle$ as a Luttinger liquid of hard-core bosons with parameter $K/\alpha$.

To compute the one-point function we must compute $S_A^{-1}(z,z')$. The solution for the ball $A=(-1, 1)$ is given by Eq. 1.19 of Ref.~\cite{Bucur_2016} up to normalization. We can use a conformal transformation to map to the general interval $A = (u, v)$, giving us
\begin{align}
\begin{split}
\label{eq:Q_A}
    &S_A^{-1}(z, z') = \\
    &-\frac{2}{K}\log \left[ \frac{(v-u)|z-z'|}{\left(\sqrt{(z-u)(v-z')} + \sqrt{(v-z)(z'-u)}\right)^2}\right]
\end{split}
\end{align}
Thus we find
\begin{equation}
\label{eq:boson_1pt}
    C^{(\alpha)}(x, A) \sim \left[\frac{(v-u)a}{(x-u)(v-x)}\right]^{\alpha/4K}
\end{equation}
Therefore, the Renyi one-point function has half the asymptotic scaling dimension as the Renyi two-point function, for example taking $x$ to lie at the midpoint of $A$ and $r=|A|$, $C^{(\alpha)}(x,A) \sim r^{-\alpha/4K}$ while $C^{(\alpha)}(x, y) \sim r^{-\alpha/2K}$.

Now we compute the entropy. Asymptotically, $S_A$ has eigenvalues $\lambda_m \sim \frac{\pi}{L}(m-\frac{1}{4})$ \cite{Kwasnicki_2010}. We use $\sum_{m=1}^{N} \log(\lambda_m) = N\log(\frac{\pi}{L}) + \log \Gamma(N+\frac{3}{4}) - \log\Gamma(\frac{3}{4}) + O(1)$ and Stirling's approximation to find
\begin{align}
\begin{split}
    S_{VN} = s_1 \frac{L}{a} + \frac{1}{8} \log (\frac{L}{a}) + s_0
\end{split}
\end{align}
Comparing with the usual CFT result $S = \frac{c}{3} \log(\frac{L}{a})$ \cite{Holzhey_1994, Vidal_2003, Calabrese_2004}, this gives an effective central charge $c_{eff} = \frac{3}{8}$ for the dephased Luttinger liquid, in contrast to $c=1$ for the pure Luttinger liquid. For fixed $|A|$ and $|C|$, the logarithmic contribution to the entropy gives a CMI which decays at long distances as
\begin{equation}
\label{eqn:cmi_long}
    I(A:C|B)\sim \frac{1}{l^2}
\end{equation}
where $l = |B|\to\infty$.

\subsection{Laughlin state}
\label{sec:laughlin}

The Laughlin wavefunction \cite{Laughlin_1983}, by mapping to a Coulomb gas, can be seen to have a Gaussian density distribution. In particular, the $\nu=1/m$ Laughlin state has $S(q) = q^2/4\pi m$ \cite{GMP_1986}, giving the two-point correlator
\begin{equation}
\label{eqn:renyi_laughlin}
    C^{(k, k')}(r)\sim\left(\frac ar\right)^{\frac{kk'}{k+k'}m}
\end{equation}
in agreement with the results of Ref.~\cite{FQHdecoherence}. 

Now we compute $S_A^{-1}$. Assuming $A$ is simply connected, we can map it to the unit disk in complex coordinates using a conformal map $f(z)$. Then we can use the solution to the Dirichlet problem on the disk to find
\begin{equation}
    S_A^{-1}(x, y) = -2m \log \left|\frac{f(x)-f(y)}{1-\overline{f(y)}f(x)} \right|
\end{equation}
and thus we find that
\begin{equation}
\label{eqn:onept_laughlin}
    C^{(\alpha)}(x,A)\sim\left(\frac{a}{R_A(x)}\right)^{\alpha m/4}
\end{equation}
where $R_A(x) = \frac{1-|f(x)|^2}{|f'(x)|}$ is the conformal radius of $A$ relative to $x$. For the entropy of a disk of radius $R$ we find
\begin{equation}
    S_{VN} = s_2 \frac{R^2}{a^2} + s_1 \frac{R}{a} + c(A) \log{\frac{R}{a}} + s_0
\end{equation}
Unlike in $d=1$, the coefficient of the logarithm is geometry dependent. In particular, $c(A)$ is the heat-kernel coefficient of the Dirichlet problem for the Laplacian on $A$, which depends on the topology of the region (it is proportional to the Euler characteristic and vanishes for an annulus), so both the logarithm and the resulting CMI are geometry dependent. For any given geometry they can be computed from the Burghelea-Friedlander-Kappeler gluing formula for determinants \cite{BFK_1992}.

\section{Infinite Dephasing: Renyi-2}
\label{sec:phases}

When the density distribution is not Gaussian (such as for Fermi liquids and generic insulators in $d\geq 2$) the XY model of Sec.~\ref{sec:xy} must be analyzed directly. In this section we do so for the Renyi-$2$ correlator, for which the kinetic term is exactly the structure factor $S(q)$ of the parent state (the Renyi-1 correlator, whose kinetic term is that of the $\sqrt p$ ensemble, is treated in Sec.~\ref{sec:diagrams}, with the same conclusions). The analysis has two parts: the spin-wave fluctuations of the ordered phase, and the vortex sector. We write the quadratic part of $S_2$ as
\begin{equation}
\label{eqn:goldstone}
    S_2[\theta] = \int \frac{d^d q}{(2\pi)^{d}}  S(q) |\theta_q|^2+\dots
\end{equation}
which is the Goldstone action of the SWSSB phase, with a bare stiffness fixed entirely by the parent state. The anharmonic terms of Eq.~\ref{eqn:S_alpha_cumulant}, i.e. the even cumulants $S^{(2k)}_c$, renormalize it. Treated perturbatively, they generate a loop expansion with propagator $(2S(q))^{-1}$ and vertices $S^{(k)}_c$, whose two-point diagrams $\Sigma(q)$ correct the kernel to $S(q)+\Sigma(q)$. This series must be applied with some caution however as there is no small parameter for the anharmonic terms, and therefore $\Sigma(q)$ receives contributions at all orders and the propagator is renormalized nonperturbatively. However, we shall use the structure of the perturbative series to argue that the renormalized kernel of each Renyi correlator has the same infrared behavior as $S(q)$. We establish the property by a power counting of the diagrams just described. Since the expansion of the Renyi-1 correlator in Sec.~\ref{sec:diagrams} is built from the same propagators and vertices, the same power counting will serve there.

\subsection{Infrared power counting}
\label{sec:power_counting}

Consider a connected diagram built from propagators $S(q)^{-1}$ on its internal lines and vertices $S^{(k)}_c(q_1,\dots,q_k)$, with $E$ external momenta $q_1,\dots,q_E$, $\sum_jq_j=0$, entering at vertices, and let $\cA(q_1,\dots,q_E)$ denote its amputated amplitude, the value of the diagram with no propagators on the external legs. Its infrared behavior is controlled by how the vertices vanish at small momenta, Eq.~\ref{eqn:Sc_charge_neutral}, and this differs between metals and insulators.

For a Fermi liquid $S(q) \sim |q|$. The cumulants $S^{(k)}_c$ are bounded and, by Eq.~\ref{eqn:Sc_charge_neutral} and the Lipschitz continuity of the Fermi-liquid cumulants, vanish at least linearly when any one argument goes to zero, uniformly in the others \footnote{This property can be proven for free fermions. For interacting Fermi liquids we expect it to hold by universality, since their equal-time cumulants have the same Fermi-surface non-analyticities}, i.e.\ $S^{(k)}_c\lesssim\kappa$ whenever at least one argument is of order $\kappa$. Consider a region of the integration in which $F$ independent linear combinations of the loop momenta are of order $\kappa$ while the remaining combinations are generic. Let $P$ be the number of internal lines whose momenta, fixed by momentum conservation, are then of order $\kappa$, and let these lines touch $V$ vertices. The region contributes $\kappa^{dF-P+V}$: $\kappa^{dF}$ from the measure, $\kappa^{-1}$ from each small line, and a single factor of $\kappa$ from each vertex touched. The $P$ small line momenta are functions of only $F$ independent small variables, so they satisfy at least $P-F$ independent linear relations, which must be supplied by momentum conservation at the $V$ vertices they touch: $V\geq P-F$. Every region therefore contributes at least $\kappa^{(d-1)F}$, and for $d\geq2$ all loop integrals converge. Moreover, when some of the external momenta are of order $\kappa$, the vertices at which they enter supply at least one factor of $\kappa$, so that $\cA\lesssim\kappa$; in particular the two-point amplitude obeys $\cA(q,-q)\lesssim|q|$, at most as singular as $S(q)$ itself.

For an insulator, exponential clustering \cite{Hastings_2006, Nachtergaele_2006} makes every $S^{(k)}_c$ analytic in all its arguments, and analyticity together with Eq.~\ref{eqn:Sc_charge_neutral} in each argument separately implies $S^{(k)}_c\lesssim\kappa^\ell$ whenever $\ell$ of its arguments are of order $\kappa$, uniformly in the remaining ones. The structure factor is analytic as well, so that at small momentum
\begin{equation}
\label{eqn:S_insulator}
    S(q)=\bar g_{ij}q_iq_j+O(q^4)
\end{equation}
where for a band insulator $\bar g_{ij}=\int_p\mathrm{Re}\,\mathrm{tr}[P_p\,\partial_iP_p\,\partial_jP_p]$ is the Brillouin-zone average of the quantum metric \cite{Provost_1980, Kohn_1964, Marzari_1997, Resta_1999, Souza_2000, Resta_2011}, $P_p$ being the projector onto the occupied bands. A small internal line now costs $\kappa^{-2}$, i.e.\ $\kappa^{-1}$ per end, and every end of a small line is compensated by the factor $\kappa$ it supplies at its vertex. A region with $F$ small loop combinations therefore contributes $\kappa^{dF}$. All loop integrals converge in every dimension, the amplitudes are analytic in the external momenta, and they vanish as $\kappa^\ell$ when $\ell$ external momenta are of order $\kappa$. In particular $\cA(q,-q)=O(q^2)$, again at most as singular as $S(q)$.

Thus in both cases the renormalized kernel $S(q)+\Sigma(q)$ of the Renyi-2 XY model has the same infrared form as $S(q)$, $\propto|q|$ for a Fermi liquid and $\propto q^2$ for an insulator, with renormalized coefficients. To be precise, we have shown that each diagram contributing to $\Sigma(q)$ is at most as IR relevant as $S(q)$. However, as there is no small parameter in the diagrammatic series, we cannot forbid a scenario in which the dressed propagator is nonperturbatively resummed to have a different IR behavior. We do not expect such a scenario to occur in the settings considered here, as we will argue in Sec.~\ref{sec:transition}. It is easy to see that the same structure, namely the renormalized kernel having the same IR form as $S(q)$, holds for all even Renyi indices, as do the conclusions below.

\subsection{Fermi liquid}
\label{sec:phases_fl}

For a Fermi liquid, the discontinuity of the Fermi surface gives
\begin{equation}
\label{eqn:Sq_metal}
    S(q) = O(|q|)
\end{equation}
at small $q$, and by Sec.~\ref{sec:power_counting} the renormalized kernel has the same form. The Goldstone fluctuations are then
\begin{equation}
    \langle \theta^2 \rangle \sim \int \frac{d^d q}{(2\pi)^{d}} \frac{1}{S(q)+\Sigma(q)} \sim \int \frac{d^dq}{(2\pi)^d}\frac{1}{|q|}
\end{equation}
which is IR finite for $d \geq 2$. Thus in these dimensions the ordered phase is self-consistent, and the Goldstone modes give $O(r^{-(d-1)})$ subleading corrections.

The nonlocal kinetic term also controls the vortex sector. With a $|q|$ kernel, a vortex configuration of linear dimension $r$ costs an action $\propto r^{d-1}$. On the other hand, the entropy of such a configuration is $\propto \log r$ in $d=2$ and $\propto r^{d-2}$ for $d\geq3$. The action is therefore parametrically larger, forbidding macroscopic proliferation of vortices. Therefore the SWSSB phase of dephased metals is stable against both spin waves and vortices. The only mechanism by which the XY model can be disordered is if the renormalized stiffness, i.e.\ the coefficient of $|q|$ in $S(q)+\Sigma(q)$, vanishes. If this is the case, then the correlator is instead power-law, scaling as the couplings at long distances, i.e.\ as $\int_q |q| e^{iq\cdot r} \sim r^{-(d+1)}$, in accordance with the paramagnetic phase of the long-range XY model \cite{Fisher_1972_LongRange, Spohn_1999, Gross_1979, Ginibre_1970}. Therefore we have
\begin{equation}
C^{(2)}(r) = 
\begin{cases}
    C^{(2)}(\infty) + O(r^{-(d-1)}), &\text{ferromagnet phase} \\
    A r^{-(d+1)}, &\text{paramagnet phase}
\end{cases}
\end{equation}

The mean-field phase diagram however suggests that $\rho_s$ remains finite at $t=\infty$ (Sec.~\ref{sec:transition}) indicating the ferromagnetic phase is realized, and this prediction has been verified numerically \cite{Kaixiang_SpinLiquid} and experimentally \cite{Si_SWSSBObs} for the free Fermi sea. It is also supported by a previously derived inequality \cite{Abhi_Inequality} showing that the SWSSB order parameter is the dominant correlator in the doubled state. The paramagnetic scenario is also forbidden for the Renyi-1 correlator, which is the true probe of SWSSB, by the inequality Eq.~\ref{eqn:C1_bound}, as we shall argue in Sec.~\ref{sec:renyi1_phases}. 

\subsection{Insulator}
\label{sec:phases_ins}

For an insulating parent state, the Goldstone fluctuations $\int_q[S(q)+\Sigma(q)]^{-1}\sim\int_q q^{-2}$ are IR finite for $d \geq 3$, giving $O(r^{-(d-2)})$ subleading corrections in the ordered phase, and logarithmically divergent in $d=2$. Unlike the metallic case, the local kernel means that vortices can proliferate at finite stiffness. A vortex configuration costs an action $\propto\log r$ in $d=2$ and $\propto r^{d-2}$ in $d\geq 3$, scaling the same as the configurational entropy of the defect. Thus unlike the metallic case, vortices can proliferate at a critical renormalized stiffness, destroying the order. This is the familiar BKT transition in $d=2$ \cite{Berezinskii_1971, Kosterlitz_1973} and 3D XY universality in $d=3$ \cite{Chaikin_Lubensky}. The particular phase realized is determined by the full renormalized stiffness (the coefficient of $q^2$ in $S(q)+\Sigma(q)$) which is nonperturbative. Therefore for $d\geq 3$ we have
\begin{equation}
C^{(2)}(r) = 
\begin{cases}
    C^{(2)}(\infty) + O(r^{-(d-2)}) & \text{ferromagnet phase}\\
     Ae^{-r/\xi} & \text{paramagnet phase}
\end{cases}
\end{equation}
Note that the correlation length $\xi$ need not coincide with the correlation length of the parent state. The bare stiffness is $K_2^{\rm bare}=2\bar{g}$. Thus we expect that the ferromagnet (SWSSB) phase is realized when $\bar{g}$ is sufficiently large. Since $\bar{g}$ grows with the correlation length and diverges at a metallic transition, we therefore expect the SWSSB phase to manifest when the parent state is sufficiently close to a metallic transition, while deep in the insulating phase the XY model will be in the paramagnet phase and no SWSSB occurs. 

In the marginal case of $d=2$ we instead have quasi-long-range order in the ferromagnet phase,
\begin{equation}
C^{(2)}(r) = 
\begin{cases}
     A \left(\frac{a}{r}\right)^{\eta} & \text{ferromagnet phase}\\
     Ae^{-r/\xi} & \text{paramagnet phase}
\end{cases}
\end{equation}
Irrelevance of vortices requires the renormalized stiffness to exceed the universal Berezinskii-Kosterlitz-Thouless value, $K_R>2/\pi$ \cite{Nelson_1977}, i.e. $\eta=1/(2\pi K_R)<1/4$. A two-dimensional insulator can thus display quasi-long-range Renyi-2 SWSSB at infinite dephasing only with a sufficiently small exponent, $\eta<1/4$. Generically we expect the renormalized stiffness to be smaller than the bare stiffness, $K_R < K_2^{\rm bare}$, which we prove for the Renyi-1 XY model in Sec.~\ref{sec:diagrams}. This gives the necessary condition 
\begin{equation}
\label{eqn:g_bound_renyi2}
    \bar g > 1/\pi
\end{equation} for the Renyi-2 correlator to be quasi-long-range.

In fact, this bound can be related to the energy gap $\delta E$. In particular, since every excited state has energy $\omega \geq \delta E$, we have
\begin{align}
\begin{split}
\label{eq:Sq_ineq}
    S(q) \leq \frac{1}{\delta E} \int_{0}^{\infty} d\omega\,  \omega S(q, \omega) &= \frac{1}{2\delta E} \langle \psi_0| [n_{-q}, [H, n_q]] | \psi_0 \rangle
\end{split}
\end{align}
where $S(q,\omega)$ is the dynamical structure factor. The above inequality follows from $S(q) = \int_{0}^{\infty} d\omega \, S(q, \omega)$ and expanding $S(q,\omega)$ in the energy eigenbasis \cite{Feynman_1954, GMP_1986}. Assuming the parent Hamiltonian takes a tight-binding form $H = H_{\text{kin}} + H_{\text{int}}$ where $H_{int}$ is diagonal in the occupation basis and $H_{kin} = \sum_r \sum_{a} t_{a} c^\dagger_{r} c_{r+a}$, we find
\begin{align}
\begin{split}
\label{eqn:K_der}
    &\frac{1}{2} \langle \psi_0| [n_{-q}, [H, n_q]] | \psi_0 \rangle = \\&-\frac{1}{N_s} \sum_{r,a} (1-\cos(q\cdot a)) t_a \langle \psi_0 | c_r^\dagger c_{r+a} | \psi_0 \rangle
\end{split}
\end{align}
Thus, defining the kinetic energy tensor
\begin{equation}
    \cK_{ij} = -\frac{1}{N_s}\sum_r \sum_{a} a_i a_j t_{a} \langle \psi_0| c^\dagger_r c_{r+a}|\psi_0\rangle
\end{equation}
and expanding Eq.~\ref{eqn:K_der} to second order in $q$, we find
\begin{equation}
\label{eqn:g_bound}
    \bar{g}_{ij} \leq \frac{\cK_{ij}}{2\delta E}
\end{equation}
Therefore, quasi-long-range order in $d=2$ is only possible if
\begin{equation}
\label{eqn:K_bound_renyi2}
\pi \cK \geq 2\delta E
\end{equation}
for a trivial insulator.

\subsection{Quantum Hall insulators}
\label{sec:hall}

Insulators with a nonzero Hall conductance $\sigma_{xy}$ are exempt from the threshold just described. This is because vortices of quantum Hall insulators carry charge $\sigma_{xy}$ under the weak local symmetry of Eq.~\ref{eqn:weak_local} \cite{Abhi_Gutz}. This follows from the Hall response of the doubled (Choi) representation of the dephased state, and we defer the proof to Sec.~\ref{sec:hall_replica}, where it is given for every even Renyi index. A single vortex is therefore forbidden, and a vortex-antivortex pair, which carries charges $\pm\sigma_{xy}$ at its two cores, can appear only if the parent state's own fluctuations transport the charge $\sigma_{xy}$ across the separation $r$, an amplitude that decays as $e^{-r/\xi}$ in a gapped state. Vortices are therefore linearly confined at any stiffness like the metallic case and cannot proliferate. Quantum Hall insulators therefore display quasi-long-range SWSSB in $d=2$ at infinite dephasing generically. This justifies the treatment of Laughlin states using the Gaussian description of Sec.~\ref{sec:laughlin}.

\section{Finite Dephasing and the Doubled State}
\label{sec:replica}

The XY representation is a statement about infinite dephasing. In this section we address three questions that lie outside it. First, whether the SWSSB found at $t=\infty$ survives to large but finite $t$. Second, the nature of the transition at which SWSSB sets in as $t$ increases from zero. Third, the statement of Sec.~\ref{sec:hall} that the vortices of a quantum Hall insulator carry the weak charge. We answer the latter two questions by mapping the doubled (Choi) representation of the dephased state \cite{Choi_Isomorphism, Jamiolkowski_Isomorphism, Lee_2025, Kaixiang_SpinLiquid, Bao_2026} to a replica field theory \cite{Abhi_Gutz}. We work with the Renyi-2 correlator, though the results again generalize straightforwardly to other even Renyi indices.

\subsection{Finite dephasing as a symmetry-breaking field}
\label{sec:finite_t}

The perturbation of the $t=\infty$ ensemble by finite $t$ is easily identified by expanding in off-diagonal elements \cite{FQHdecoherence}. Off-diagonal elements of $\rho$ in which ket and bra differ by moving one fermion from $z$ to $w$, which we refer to as charge-transfer defects, carry a factor $e^{-t}$, giving
\begin{align}
\begin{split}
\label{eqn:xy_field}
\tr\rho^2&=\sum_{n,n'}p[n]\,p[n']\,e^{-t\sum_z(n_z-n'_z)^2}\\
&=\int D\theta\,e^{-S_2[\theta]}\prod_z\left(1+2e^{-t}\cos\theta_z\right)
\end{split}
\end{align}
so that large but finite dephasing corresponds to a weak pinning field $\sum_z 2e^{-t} \cos(\theta_z)$ acting on the XY model, breaking the weak local symmetry Eq.~\ref{eqn:weak_local} explicitly.

However, the Renyi-2 correlator at finite $t$ is not the spin-spin correlator of the XY model with the pinning field. We can also expand the numerator of the Renyi-2 correlator about $t=\infty$ to find
\begin{align}
\begin{split}
\label{eqn:numerator_finite_t}
\tr[\rho c^\dagger_xc_y\rho c^\dagger_yc_x]&=\sum_{n,n'}b[n]\,b^*[n']\,e^{-t\sum_z(n_z-n'_z)^2}\\
b[n]&=\langle\psi_0|c_y^\dagger c_x |n\rangle\langle n|\psi_0\rangle
\end{split}
\end{align}
and the Fourier representation of Eq.~\ref{eqn:xy_field} gives
\begin{align}
\begin{split}
\label{eqn:C2_finite_t}
C^{(2)}_t(x,y)&=\frac{\int D\theta\,e^{-S_2[\theta]}\,|\cG_{xy}[\theta]|^2\prod_z\left(1+2e^{-t}\cos\theta_z\right)}{\int D\theta\,e^{-S_2[\theta]}\prod_z\left(1+2e^{-t}\cos\theta_z\right)}\\
\cG_{xy}[\theta]&=\frac{\langle\psi_0|c^\dagger_yc_x\,e^{i\theta\cdot n}|\psi_0\rangle}{\langle\psi_0|e^{i\theta\cdot n}|\psi_0\rangle}
\end{split}
\end{align}

The correct finite-$t$ insertion is therefore $|\cG_{xy}|^2$. At $t=\infty$ the insertions $|\cG_{xy}|^2$ and $e^{i(\theta_x-\theta_y)}$ have the same average, both integrating to $\sum_np[n]p[n+h]$, and either may be used, but at finite $t$ they respond oppositely to the pinning field. At the pinned configuration $\theta=0$, $\cG_{xy}[0]=G(x-y)$ is the parent correlation function, so pinning drives $C^{(2)}$ toward its undephased value $|G(x-y)|^2$, which decays, whereas the spin-spin correlator would saturate. 

The fate of the SWSSB found at $t=\infty$ then follows from the response of the XY model to a weak pinning field, and it differs between the phases of Sec.~\ref{sec:phases}. In a long-range ordered phase, dephased metals in $d\geq2$ and insulators above threshold in $d\geq3$, a weak field fixes the zero mode of $\theta$ and gaps the Goldstone modes beyond a length that diverges as $t\to\infty$. Since $|\cG_{xy}[\theta]|^2$ is invariant under global shifts of $\theta$, and the Goldstone fluctuations are infrared finite, removing them beyond that length changes $C^{(2)}$ by an amount that vanishes as $t\to\infty$. Therefore the long-range order is stable to a sufficiently weak pinning field, and it persists at large but finite $t$ down to a critical threshold $t_c$ whose critical theory is a Ginzburg-Landau theory as explained in Sec.~\ref{sec:transition}. In the quasi-long-range ordered phase of insulators in $d=2$, the pinning field may be relevant or irrelevant. Since $\langle e^{i(\theta_x-\theta_y)}\rangle\sim r^{-\eta}$ at the $t=\infty$ fixed point, the scaling dimension of the pinning field $\cos\theta_z$ there is $\eta/2$, so the field is relevant when $\eta<4$. Any finite $t$ then pins $\theta$ beyond a length $\xi_p\propto e^{2t/(4-\eta)}$, the dephased state is coherent at longer scales, and by Eq.~\ref{eqn:C2_finite_t} the Renyi-2 correlator reverts beyond $\xi_p$ to the decay of $|G|^2$, which is short-ranged for an insulator. For $\eta>4$ it is irrelevant, the quasi-long-range order survives to a finite $t_c$, and a Berezinskii-Kosterlitz-Thouless transition takes place there \cite{FQHdecoherence}. A trivial two-dimensional insulator can only be quasi-long-range ordered with $\eta<1/4<4$ (Sec.~\ref{sec:phases_ins}), so its SWSSB at infinite dephasing is destroyed by any finite amount of coherence. For Laughlin states $\eta=m$, so the order is destroyed for $m<4$, as verified numerically for $m=1$ \cite{Abhi_CFTChern}, and survives for $m>4$ \cite{FQHdecoherence}. The vortex language of Sec.~\ref{sec:hall} reproduces this criterion - the plasma analogy makes the Gaussian kernel of the Laughlin state exact, so the elementary vortex has dimension $\pi K=1/2m$, and the electron-transfer defect, an $m$-fold vortex, has dimension $m^2/2m=m/2=\eta/2$. We expect the same conclusions to hold for the Renyi-1 correlator, although the expansion of Sec.~\ref{sec:diagrams} is only defined at $t=\infty$.

\subsection{Replica field theory from the doubled state representation}
\label{sec:transition}

At $t=0$ the state is pure and the Renyi-2 correlator reduces to the parent correlation function, $C^{(2)}(x,y)=|\langle\psi_0|c^\dagger_xc_y|\psi_0\rangle|^2=|G(x-y)|^2$, which decays: the strong symmetry is unbroken. Wherever the SWSSB found at $t=\infty$ persists to finite $t$ (Sec.~\ref{sec:finite_t}) a transition therefore takes place at some finite $t_c$, and we now identify its universality class by means of a replica field theory of the doubled state. Under the Choi isomorphism \cite{Choi_Isomorphism, Jamiolkowski_Isomorphism} the density matrix maps to the doubled state \cite{Lee_2025, Kaixiang_SpinLiquid, Bao_2026}
\begin{equation}
\label{eqn:choi_dephase}
|\rho\rangle\rangle = e^{-\frac{t}{2} \sum_i (n_{i,L} - n_{i,R})^2}|\psi_{0,L}\rangle |\psi^*_{0,R}\rangle
\end{equation}
In this language $\tr\rho^2=\langle\langle\rho|\rho\rangle\rangle$, and the Renyi-2 correlator is the two-point function of the interlayer pair $\Delta_z=c_{z,R}c_{z,L}$ in the doubled space,
\begin{equation}
\label{eqn:C2_choi}
C^{(2)}(x,y)=\frac{\langle\langle\rho|\Delta^\dagger_x\Delta_y|\rho\rangle\rangle}{\langle\langle\rho|\rho\rangle\rangle}
\end{equation}
The pair carries charge two under the strong $U(1)$ and none under the weak $U(1)$, and at $t=\infty$ its phase is the spin $e^{i\theta_z}$ of Sec.~\ref{sec:xy}. More generally, for every even Renyi index the correlator is the same two-point function in the doubled state of $\rho^{\alpha/2}$,
\begin{align}
\begin{split}
\label{eqn:choi_alpha}
C^{(\alpha)}(x,y)&=\frac{\langle\langle\rho^{\alpha/2}|\Delta^\dagger_x\Delta_y|\rho^{\alpha/2}\rangle\rangle}{\langle\langle\rho^{\alpha/2}|\rho^{\alpha/2}\rangle\rangle}\\
|\rho^{\alpha/2}\rangle\rangle&=\rho_L^{\alpha/2-1}|\rho\rangle\rangle
\end{split}
\end{align}
where $\rho_L=\rho\otimes1$: the additional factors of $\rho$ are copies of the parent state inserted on the $L$ layer, and at $t=\infty$, where every factor is diagonal, $|\rho^{\alpha/2}\rangle\rangle=\sum_np[n]^{\alpha/2}|n\rangle_L|n\rangle_R$. SWSSB is long-range order of $\Delta$, i.e.\ its condensation in the doubled state, and the transition at $t_c$ is the onset of this condensation.

Eq.~\ref{eqn:choi_dephase} defines a field theory of two fermion flavors with a decoherence-induced attractive interaction at $\tau=0$ coupling the flavors \cite{Lee_2023, Kaixiang_CFT, Abhi_CFTChern}. In fact, since both $|\rho\rangle\rangle$ and $\langle\langle\rho|$ carry the decoherence factor, the continuum interaction is supported on a thin temporal slab $|\tau|\leq\epsilon$ of vanishing width, and this regularization is essential for the continuum theory to reproduce the correct physics \cite{Abhi_Gutz}. The interaction is decoupled in the pairing channel by a Hubbard-Stratonovich transformation \cite{Stratonovich_1957, Hubbard_1959},
\begin{align}
\begin{split}
    e^{-\frac{t}{2} \sum_i (n_{i,L} - n_{i,R})^2} \sim \prod_i \int d^2\phi_i\, &e^{-\frac{1}{t} |\phi_i|^2}\\
    &\times e^{-[\phi_i c_{i,R} c_{i,L} + h.c.]}
\end{split}
\end{align}
With $\phi_\pm$ the fields from $|\rho\rangle\rangle$ and $\langle\langle\rho|$, $\phi_a=\phi_++\phi_-$ and $\phi_r=\epsilon(\phi_+-\phi_-)$, the continuum theory is
\begin{align}
\begin{split}
    &S = S_0[c_L] + \bar{S}_0[c_R] + \frac{1}{2\epsilon} \int_{-\epsilon}^{\epsilon}d\tau \int d^dx \cL_{\text{dec}}\\
    \cL_{\text{dec}} &= \frac{1}{2t}[|\phi_a(x)|^2 + |\phi_r(x)|^2/\epsilon^2] \\
    &-[(\phi_a(x) + \phi_r(x) \partial_\tau)\Delta(x,\tau) + h.c.]
\end{split}
\end{align}
with $S_0$ the bulk action preparing $|\psi_0\rangle$ and $\bar S_0$ its time reversal; the normalizations make $\phi_{a,r}$ order-one fields as $\epsilon\to0$ \cite{Abhi_Gutz}. The equations of motion give $\phi_a(x)=t\Delta(x,0)$, so the Renyi-2 correlator is the two-point function of $\phi_a$ along $\tau=0$, $C^{(2)}(r)\sim\langle\phi_a^*(0)\phi_a(r)\rangle$, and long-range order of $\phi_a$ is SWSSB. Integrating out the fermions at quadratic order,
\begin{equation}
\label{eqn:S_eff_phi}
    S_{eff}[\phi_a] = \frac{1}{2}\int \frac{d^d q}{(2\pi)^{d}} \left[\frac{1}{2t} - 2\chi_\Delta(q)\right] |\phi_a(q)|^2
\end{equation}
where
\begin{equation}
\label{eqn:chi_def}
    \chi_\Delta(q) = \frac{1}{(2\epsilon)^2}\int_{-\epsilon}^{\epsilon} d\tau_1 d\tau_2 \langle \Delta^*(-q,\tau_1) \Delta(q, \tau_2)\rangle_0
\end{equation}
is the regulated equal-time Cooper pair susceptibility of the undephased doubled state $|\psi_{0,L}\rangle|\psi^*_{0,R}\rangle$. Since this state is a product of the two layers, the pair correlators factorize, $\langle\Delta^\dagger_x\Delta_y\rangle_0=\langle c^\dagger_{x,L}c_{y,L}\rangle\langle c^\dagger_{x,R}c_{y,R}\rangle=|G(x-y)|^2$, and $\chi_\Delta$ is determined for any parent state by its momentum occupation function $f(p)=\langle\psi_0|c^\dagger_pc_p|\psi_0\rangle$,
\begin{align}
\begin{split}
\label{eqn:chi_eval}
    \chi_\Delta(q) &= \tfrac12\int_p\left[(1-f(p))(1-f(q-p)) + f(p)f(q-p)\right] \\
    &= \chi_\Delta(0) - S_{HF}(q)
\end{split}
\end{align}
where $\int_p\equiv\int\frac{d^dp}{(2\pi)^d}$, $\chi_{\Delta}(0) = \frac{1}{2}-\nu + \int_p f(p)^2 > 0$ and
\begin{equation}
\label{eq:hf_def}
    S_{HF}(q) = \int_p f(p) \left[f(p) - f(p+q)\right]
\end{equation}
is the Hartree-Fock equal-time structure factor. For free fermions $S_{HF}(q) = S(q)$ exactly.

Eqs.~\ref{eqn:S_eff_phi}--\ref{eq:hf_def} give a Ginzburg-Landau theory for the SWSSB order parameter,
\begin{align}
\begin{split}
\label{eqn:GL}
    S_{eff}[\phi_a] &= \int_q \left[\mu(t) + S_{HF}(q)\right]|\phi_a(q)|^2 \\&\quad+ u\int d^dx\,|\phi_a|^4+\dots \\
    \mu(t)&=\frac{1}{4t}-\chi_\Delta(0)
\end{split}
\end{align}
The mass vanishes at $t_c^{MF}=1/4\chi_\Delta(0)$, and for $t>t_c^{MF}$ the mean-field saddle point $\phi_a\ne0$ is ordered. Writing $\phi_a=\sqrt{\rho_s+\delta\rho}\,e^{i\theta}$ in the ordered phase, with $\rho_s(t)\sim 1-t_c/t$ near the transition, the amplitude $\delta\rho$ is gapped and the phase has the Goldstone action $\rho_s\int_q S_{HF}(q)|\theta_q|^2$ plus higher order terms. As $t\to\infty$ the Goldstone action becomes the exact XY action $S_2[\theta]$ of Eq.~\ref{eqn:S_alpha_cumulant}. The SWSSB phase of Sec.~\ref{sec:phases} is thus the large-$t$ end of the ordered phase of Eq.~\ref{eqn:GL}.

Near the transition at $t_c$ the replica theory of a Fermi liquid therefore takes the universal real-space form
\begin{equation}
\label{eqn:GL_metal}
S_{eff}[\phi_a]=\int d^dx\;\phi_a^*|\partial|\phi_a+\mu|\phi_a|^2+u|\phi_a|^4
\end{equation}
where the fractional Laplacian $|\partial|\leftrightarrow|q|$ captures the nonlocal effect of the gapless bulk. The upper critical dimension is $d_c=2$ \cite{Fisher_1972_LongRange}, so the transition is mean-field in $d>2$ and marginal in $d=2$. For an insulator the kinetic term is the usual $|\partial\phi_a|^2$ with $d_c=4$, and the transition into the long-range ordered phase in $d=3$ is in the 3D XY universality class. In $d=2$ a finite-$t$ transition into a quasi-long-range ordered phase exists only for $\eta>4$ (Sec.~\ref{sec:finite_t}), which is described by the Berezinskii-Kosterlitz-Thouless universality class \cite{FQHdecoherence}. We expect the same universality classes for every even Renyi index, since by Eq.~\ref{eqn:choi_alpha} $C^{(\alpha)}$ is the two-point function of the same pair in a doubled state whose undephased limit is again $|\psi_{0,L}\rangle|\psi^*_{0,R}\rangle$, so that the singularity of the kernel is the same. 

This description also gives a physical reason why we do not expect a nonperturbative renormalization of the kernel to alter its infrared form even at $t=\infty$. In the representation of Eq.~\ref{eqn:choi_dephase} the decoherence acts on the doubled ground state only on the slab around $\tau=0$, and cannot open a gap in the bulk. For a metal the $|q|$ singularity of $S_{HF}(q)$ is the Fermi-surface singularity of the bulk fermions, which is present even in the nonperturbative regime $t=\infty$. What a resummation can do is renormalize the coefficient of $|q|$, i.e.\ $\rho_s$, and the phase transition corresponds to $\rho_s\to0$ at $t_c$. That $\rho_s$ remains finite at $t=\infty$ is the mean-field prediction, confirmed by numerics and experiment in $d=2$, but it is not established by a controlled calculation. For an insulator all ingredients are analytic, and a resummation can produce a nonanalyticity only at a phase transition, which in the phase-only language is the vortex transition of Sec.~\ref{sec:phases_ins}.

\subsection{Vortices of quantum Hall insulators}
\label{sec:hall_replica}

The doubled representation also proves the statement of Sec.~\ref{sec:hall} that the vortices of a quantum Hall insulator carry the weak charge. We give the argument for the Renyi-2 correlator, which extends directly to every even Renyi index. The order parameter $\phi_a\sim\Delta=c_Rc_L$ carries strong charge two and no weak charge, and its phase is the XY variable $\theta$ of Sec.~\ref{sec:xy}, $e^{i\theta_z}$ being the phase of the interlayer pair $c_{z,L}c_{z,R}$. We couple the two layers to background gauge fields $A_L$ and $A_R$ and write $A_{L,R}=A_+\pm A_-$, so that $A_+$ couples to the strong charge $n_L+n_R$ and $A_-$ to the weak charge $n_L-n_R$. The Chern-Simons response of the parent state and of its conjugate copy,
\begin{equation}
    \frac{\sigma_{xy}}{4\pi}\left(A_L dA_L - A_R dA_R\right) = \frac{\sigma_{xy}}{\pi}\, A_+ dA_-
\end{equation}
is a mixed term. In the SWSSB phase gauge invariance requires $A_+$ to appear in the combination $A_+-\frac12\partial\theta$, and varying the mixed term with respect to $A_{-,0}$ gives the weak charge density bound to the phase field,
\begin{equation}
\label{eqn:flux_attachment}
    \rho_- = \frac{\sigma_{xy}}{\pi}\,B_+ - \frac{\sigma_{xy}}{2\pi}\,\epsilon^{ij}\partial_i\partial_j\theta
\end{equation}
With no background field, a $2\pi$ vortex, $\epsilon^{ij}\partial_i\partial_j\theta=2\pi\delta^2(x-x_v)$, carries weak charge $\sigma_{xy}$. Conversely a unit weak charge, i.e. a ket-bra mismatch of $\rho$, which is precisely the charge-transfer defect of Sec.~\ref{sec:finite_t}, is a vortex of winding $2\pi/\sigma_{xy}$. For a $\nu=1/m$ Laughlin state the elementary vortex is a quasiparticle-transfer defect of weak charge $1/m$, and the electron-transfer defect is an $m$-fold vortex. In a trivial insulator the two are independent, i.e.\ the vortex is weak-neutral and the defect carries no vorticity. By Eq.~\ref{eqn:choi_alpha} the same pair and the same two layers carry the argument for every even Renyi index, so that vortices are confined at $t=\infty$ for all of them. We expect this to extend to the Renyi-1 XY model by continuity in the Renyi index.

\section{Infinite Dephasing: Renyi-1}
\label{sec:diagrams}

The Renyi-1 correlator is the spin-spin correlator of the XY model of Sec.~\ref{sec:xy} with weight $|\Psi_1|^2$, the squared full counting statistics of the ensemble $\sqrt p$. Unlike for the Renyi-2 correlator, the couplings of this model are not correlation functions of the parent state. The cumulants $\kappa^{(1)}_k$ of Eq.~\ref{eqn:S_alpha_cumulant} are those of $\sqrt p$, and it is unclear how to compute them a priori. In this section we introduce a diagrammatic expansion that expresses
\begin{equation}
\label{eqn:F_h_def}
F[h]\equiv-\log\frac{\cB_1[h]}{\cB_1[0]}=-\log\sum_n\sqrt{p[n]\,p[n+h]}
\end{equation}
for an arbitrary integer charge-transfer configuration $h$ in terms of the ordinary connected correlation functions $S^{(k)}_c$ of the parent state. The two-point function is the case of a single dipole, $h=\delta_x-\delta_y$, for which $C^{(1)}(x,y)=e^{-F[h]}$. Organized by the number of factors of $h$,
\begin{equation}
\label{eqn:F_h_kernels}
    F[h]=\frac18\int_q K_2(q)|h_q|^2+\frac{1}{4!}\sum K_4\,h^4+\dots
\end{equation}
with $h_q=\sum_zh_ze^{-iq\cdot z}$, where $\int_q$ runs over the Brillouin zone and $\sum K_kh^k$ is shorthand for the contraction of the kernel $K_k(z_1,\dots,z_k)$ with $k$ factors of $h$. This is precisely the spin-wave expansion of the Renyi-1 XY model in the charge representation of Sec.~\ref{sec:spinwave}. The expansion of $F[h]$ in powers of $h$ is the Fourier dual of the expansion of $S_1[\theta]$ in powers of $\theta$. The two are related by a Legendre transform at tree level and by loop corrections beyond it, with propagator $4K_2^{-1}$ and vertices $K_4,K_6,\dots$, so that the kinetic term of the Renyi-1 XY model is
\begin{equation}
\label{eqn:kappa1_K2}
\kappa^{(1)}_2(q)=2K_2(q)^{-1}+\dots
\end{equation}
and the higher $K_{2j}$ determine its anharmonic couplings. The expansion applies to any classical distribution, i.e.\ any diagonal ensemble, and can be used outside the setting of decohered ground states to study any classical data.

The expansion is built on the likelihood ratio of the shifted distribution,
\begin{equation}
\label{eqn:lr_def}
R_h[n]=\frac{p[n+h]}{p[n]}
\end{equation}
where $p$ vanishes outside the space of occupation configurations, and $\langle\cdot\rangle$ denotes an average over $p[n]$, i.e.\ an expectation value in $|\psi_0\rangle$. Since $n\to n+h$ merely relabels the configurations, $R_h$ satisfies the exact identity
\begin{equation}
\label{eqn:ibp}
\langle R_h\,F\rangle=\langle T_{-h}F\rangle,\qquad (T_{-h}F)[n]=F[n-h]
\end{equation}
for any function $F$ of the configuration; in particular $\langle R_h\rangle=1$. In terms of the discrete score
\begin{equation}
\label{eqn:score}
\sigma_h[n]=1-R_h[n]
\end{equation}
which has zero mean, Eq.~\ref{eqn:ibp} reads $\langle\sigma_hF\rangle=\langle\nabla_hF\rangle$ with $\nabla_hF=F-T_{-h}F$ the finite difference along $h$. This is the lattice form of Stein's identity \cite{Stein_1981}: in the continuum limit $\sigma_h$ becomes the score $\sum_zh_z\,\partial\Gamma/\partial n_z$ of $\Gamma=-\log p$ \cite{Hyvarinen_2005} and $\nabla_h$ the directional derivative. Eq.~\ref{eqn:F_h_def} is the average of $\sqrt{R_h}$, so that
\begin{align}
\begin{split}
\label{eqn:F_moments}
F[h]&=-\log\left\langle\sqrt{1-\sigma_h}\right\rangle\\
&=\frac18\langle\sigma_h^2\rangle+\frac1{16}\langle\sigma_h^3\rangle+\frac5{128}\langle\sigma_h^4\rangle+\frac1{128}\langle\sigma_h^2\rangle^2+\dots
\end{split}
\end{align}
an expansion in the moments of the score. It is even in $h$, because $F[h]=F[-h]$ by relabeling, and it is graded by the number of factors of $h$: $\sigma_h$ is of first order in $h$, so the $k$-th moment is of order $h^k$.

The moments of $\sigma_h$ are not obviously correlation functions of $p[n]$, since $\sigma_h$ contains the ratio $p[n+h]/p[n]$. They become so once $\sigma_h$ is expanded in moments of the density, which is where Eq.~\ref{eqn:ibp} enters. We define $\delta n_z=n_z-\nu$, let $\mathcal H_m$ be the span of polynomials of degree $\le m$ in $\delta n$, $P_m$ the orthogonal projector onto $\mathcal H_m$ in the inner product $\langle f,g\rangle=\langle f^\ast g\rangle$, and
\begin{equation}
\label{eqn:Ym}
    Y_m(z_1,\dots,z_m)=(1-P_{m-1})\big[\delta n_{z_1}\cdots\delta n_{z_m}\big]
\end{equation}
the degree-$m$ product with its lower-degree components projected out, i.e.\ normal ordered so that $\langle Y_mX_{m-1}\rangle=0$ for every polynomial $X_{m-1}$ of degree $\le m-1$. Defining the overlaps and the Gram kernel
\begin{align}
\begin{split}
\label{eqn:sm_gram}
    \sigma^{(m)}(z)&=\langle \sigma_h Y_m(z)\rangle=\langle \nabla_h Y_m(z)\rangle
    \\
    \M^{(m)}(z,z')&=\langle Y_m(z)Y_m(z')\rangle
\end{split}
\end{align}
with $z=(z_1,\dots,z_m)$, the degree-$m$ part of the score is
\begin{equation}
\label{eqn:score_expansion}
    (P_m-P_{m-1})\sigma_h=\frac{1}{m!}\sum_{z,z'}Y_m(z)\big(\M^{(m)}\big)^{-1}(z;z')\,\sigma^{(m)}(z')
\end{equation}
where $(\M^{(m)})^{-1}$ acts on symmetric functions, $\sum_{z'}\M^{(m)}(z,z')(\M^{(m)})^{-1}(z';z'')=\sum_\sigma\prod_i\delta_{z_i,z''_{\sigma_i}}$. Both ingredients are correlation functions of the parent state: the Gram kernel by definition, and the overlaps because $\nabla_hY_m$ is a polynomial in $\delta n$ and $h$. The leading term of Eq.~\ref{eqn:F_moments} is then
\begin{align}
\begin{split}
\label{eqn:Delta_def}
\langle\sigma_h^2\rangle&=\sum_{m\geq1}\Delta_m\\
\Delta_m&=\big\|(P_m-P_{m-1})\sigma_h\big\|^2\\&=\frac{1}{m!}\sum_{z,z'}\sigma^{(m)}(z)^\ast\big(\M^{(m)}\big)^{-1}(z;z')\,\sigma^{(m)}(z')
\end{split}
\end{align}
and the higher moments follow by inserting Eq.~\ref{eqn:score_expansion} into $\langle\sigma_h^k\rangle$ and evaluating the overlaps of the $Y_m$.

At $m=1$, $Y_1(z)=\delta n_z$ and $\nabla_hY_1(z)=h_z$, so $\sigma^{(1)}(z)=h_z$ and $\M^{(1)}(z,z')=S(z,z')$, whence
\begin{equation}
\label{eqn:Delta1}
    \Delta_1=\sum_{z,z'}h_zS^{-1}(z,z')h_{z'}=\int_{\mathrm{BZ}}\frac{d^dq}{(2\pi)^d}\frac{|h_q|^2}{S(q)}
\end{equation}
At $m=2$,
\begin{align}
\begin{split}
    Y_{2}(z_1, z_2) &= \delta n_{z_1}\delta n_{z_2}-S(z_1,z_2)\\&-\sum_{w,w'}\delta n_wS^{-1}(w,w')S^{(3)}_c(w',z_1,z_2)
\end{split}
\end{align}
and the finite difference gives
\begin{align}
\begin{split}
\label{eqn:s2}
    \sigma^{(2)}(z_1,z_2)&=-h_{z_1}h_{z_2}\\&-\sum_{w,w'}h_wS^{-1}(w,w')S^{(3)}_c(w',z_1,z_2)
\end{split}
\end{align}
where the first term, of second order in $h$, is a finite-difference correction absent in the continuum. The Gram kernel is
\begin{align}
\begin{split}
&\M^{(2)}(z_1,z_2;z_1',z_2')=S(z_1,z_1')\,S(z_2,z_2')\\
&\quad+S(z_1,z_2')\,S(z_2,z_1')+S^{(4)}_c(z_1,z_2,z_1',z_2')\\
&\quad-\sum_{w,w'}
S^{(3)}_c(z_1,z_2,w)\,S^{-1}(w,w')\,S^{(3)}_c(w',z_1',z_2')
\end{split}
\end{align}

The process continues to all $m$. Expanding each Gram inverse $(\M^{(m)})^{-1}$ in a geometric series about its Gaussian part $\M^{(m)}_{\rm gauss}=\sum_\sigma\prod_iS(z_i,z'_{\sigma_i})$, every term of Eq.~\ref{eqn:F_moments} becomes a finite diagram built from
\begin{itemize}
\itemsep0pt
\item propagators $S^{-1}$
\item external legs $h$
\item vertices $S^{(k)}_c$ with $k\geq3$ (two-point factors $S$ in the Gram kernels always sit between two propagators and simply merge them)
\end{itemize}
Disconnected diagrams cancel between the moments of Eq.~\ref{eqn:F_moments}, as they must in a logarithm, and $F[h]$ is the sum of the connected ones. When $p[n]$ is Gaussian there are no vertices, the only connected diagram is $\Delta_1$, and
\begin{equation}
\label{eqn:quad_c1_exact}
    F[h]=\tfrac18\Delta_1
\end{equation}
exactly, reproducing Eq.~\ref{eq:renyi_1d} at $k=k'=1$. This is a nontrivial check on Eqs.~\ref{eqn:F_moments}--\ref{eqn:Delta1}. 

\subsection{Phases}
\label{sec:renyi1_phases}

In momentum space every diagram contributing to $F[h]$ consists of $E$ external legs $h_{q_j}/S(q_j)$, $\sum_jq_j=0$, attached to an amputated amplitude $\cA(q_1,\dots,q_E)$ that collects the vertices $S^{(k)}_c$, the internal propagators $S^{-1}$ and the loop integrations. These are precisely the amplitudes analyzed in Sec.~\ref{sec:power_counting}, and the power counting of that section applies to them verbatim. In particular the two-leg diagrams sum to $\frac18\int_q|h_q|^2K_2(q)$ with $K_2=S^{-1}+S^{-1}\cA S^{-1}+\dots$, and since $\cA(q,-q)\lesssim|q|$ for a Fermi liquid and $\cA(q,-q)=O(q^2)$ for an insulator, every correction to $K_2(q)$ is at most as singular as $S(q)^{-1}$ itself. By Eq.~\ref{eqn:kappa1_K2} the renormalized kernel of the Renyi-1 XY model, i.e.\ its $\theta$ two-point function, retains the infrared form of $S(q)$ as well, exactly as for the Renyi-2 model. Its bare value is $2S(q)$, by Eq.~\ref{eqn:quad_c1_exact}.

All conclusions of Sec.~\ref{sec:phases} therefore carry over to the Renyi-1 correlator. However, for the Renyi-1 case the paramagnetic scenario of the long-range XY model, in which $C^{(1)}(r)\sim r^{-(d+1)}$ would decay faster than the parent correlator $|G(r)|\sim r^{-(d+1)/2}$, violates the bound Eq.~\ref{eqn:C1_bound}. Therefore we conclude that dephased metals have long-range Renyi-1 SWSSB in $d\geq2$, \begin{equation}
    C^{(1)}(r)=C^{(1)}(\infty)+O(r^{-(d-1)})
\end{equation}
with a nonuniversal plateau $C^{(1)}(\infty)$ to which every diagram contributes. This has been confirmed numerically \cite{Kaixiang_SpinLiquid} and experimentally \cite{Si_SWSSBObs} in $d=2$. The phase diagram of the dephased insulators carries over verbatim from the Renyi-2 case, so that in $d\geq3$
\begin{equation}
C^{(1)}(r) = 
\begin{cases}
    C^{(1)}(\infty) + O(r^{-(d-2)}) & \text{ferromagnet phase}\\
     Ae^{-r/\xi} & \text{paramagnet phase}
\end{cases}
\end{equation}
while in $d=2$ 
\begin{equation}
C^{(1)}(r) = 
\begin{cases}
     A\left(\frac ar\right)^{\eta}& \text{ferromagnet phase}\\
     Ae^{-r/\xi} & \text{paramagnet phase}
\end{cases}
\end{equation}
with $\eta<1/4$. The expansion, whose terms are bounded uniformly in $r$, cannot see the vortices, and must fail to converge when they proliferate in the disordered phase. We again expect vortices to be confined for quantum Hall insulators by continuity in the Renyi index, exempting them from the threshold. This is consistent with the exact exponents of Eq.~\ref{eqn:renyi_laughlin} at $k=k'=1$, where the Renyi-1 exponent is $m/2\geq1/2>1/4$.

The one quantitative difference is the bare stiffness, $K_1^{\rm bare}=4\bar g$, twice that of the Renyi-2 model of Sec.~\ref{sec:phases_ins}. For a two-dimensional insulator the two-leg diagrams are logarithmically divergent, $\Delta_m=c_m\log(r/a)+O(1)$, while all diagrams with more legs remain finite (Sec.~\ref{sec:power_counting}), so within the spin-wave expansion (i.e. when vortices are irrelevant) we have
\begin{equation}
\label{eqn:eta_sum}
    C^{(1)}(r)\sim\left(\frac ar\right)^{\eta},\qquad \eta=\frac18\sum_mc_m
\end{equation}
Every $\Delta_m$ is nonnegative, so $c_m\geq0$ and $\eta\geq c_1/8$, with
\begin{equation}
\label{eqn:Delta1_log}
    \Delta_1=\int\frac{d^2q}{(2\pi)^2}\frac{|h_q|^2}{\bar g_{ij}q_iq_j}\sim \frac{1}{\pi\sqrt{\det\bar g}}\log\frac ra
\end{equation}
and therefore, using Eq.~\ref{eqn:g_bound},
\begin{equation}
\label{eqn:eta_bound}
    \eta\ge\frac{1}{8\pi\sqrt{\det\bar g}}\geq\frac{\delta E}{4\pi\sqrt{\det\cK}}
\end{equation}
With $\eta=1/2\pi K_R$ the first inequality reads $K_R\leq4\sqrt{\det\bar g}=K_1^{\rm bare}$. In other words, the statement $K_R \leq K_1^{\rm bare}$ follows from the positivity of $\Delta_m$, which we expect to hold also for the Renyi-2 case as stated in Sec.~\ref{sec:phases_ins}. Combined with the BKT condition $\eta<1/4$, quasi-long-range Renyi-1 SWSSB of a two-dimensional insulator at infinite dephasing requires
\begin{equation}
\label{eqn:necessary_2d}
    \sqrt{\det\bar g}>\frac{1}{2\pi},\qquad \delta E<\pi\sqrt{\det\cK}
\end{equation}
The window is twice as wide as the Renyi-2 one of Eqs.~\ref{eqn:g_bound_renyi2} and \ref{eqn:K_bound_renyi2}, in keeping with the larger bare stiffness.

\section{Conclusions}
\label{sec:conclusions}
In this work we have studied the strong-to-weak spontaneous symmetry breaking of $U(1)$ symmetric fermions under infinite density dephasing. Our central tool is an exact representation of all Renyi correlators at infinite dephasing as spin-spin correlators of a compact XY model, whose weight is the squared full counting statistics of the classical ensemble $p^{\alpha/2}$ built from the Born probabilities of the parent state. We also study a replica field theory applicable at finite dephasing and a novel expansion for the Renyi-1 correlator of diagonal ensembles. 

We find that maximally dephased metals display long-range SWSSB in $d \geq 2$ and algebraic SWSSB in $d=1$ as diagnosed by the Renyi-1 correlator. For maximally dephased insulators, both an SWSSB and a trivial phase are possible, with the former possessing long-range Renyi correlations in $d\geq 3$ and algebraic correlations in $d=2$. The SWSSB phase is only present when the correlation length is sufficiently large. We also determine the stability of SWSSB away from infinite dephasing and discern the nature of the SWSSB transition in each case for even Renyi indices, which we expect to extend to the Renyi-1 case as well.

Several questions remain open. The perturbative methods have no small parameter at the lattice scale, so the location of the insulating thresholds, and the value of the SWSSB plateaus, are UV dependent quantitative questions that call for a direct study of the XY model of Sec.~\ref{sec:xy}. Since the bare stiffness of the Renyi-$\alpha$ XY model decreases with $\alpha$, we expect higher Renyi index correlators to disorder more easily than lower ones. Indeed, a generic mechanism to explain this phenomenon has been discussed in a recent work \cite{Yang_2026}. 

Unlike the $U(1)$ case, SWSSB of nonabelian continuous symmetries is still quite poorly understood \cite{Chong_SWSSB}. An enticing direction along the lines of this work would be to study the SWSSB properties of dephased ground states with e.g. $SU(2)$ symmetry, and Kraus operators given by $SU(2)$ singlet operators like $\vec{S}_i \cdot \vec{S}_j$ or $S_i^2$. This problem is intrinsically quantum as unlike in the $U(1)$ case even the maximally mixed state in a fixed $SU(2)$ charge sector is non-diagonal with nontrivial quantum coherences. This necessitates the development of new techniques, as our Renyi-1 expansion and our XY representation are only defined for diagonal density matrices. This problem is also intimately related to the problem of decoding of nonabelian error correction codes \cite{Wootton_2014, Brell_2014, Dauphinais_2017, Song_2025_QuantumDouble}, which is another fascinating and important direction.

\section{Acknowledgements}
We are grateful to Cenke Xu and Chong Wang for continued guidance and collaboration on related works. We also thank Matthew Fisher, Thomas Kiely, Parsa Pourghasem, Yimu Bao, Ruochen Ma, and Leo Lessa for helpful discussions. We dedicate this work to the 2026 New York Knicks.

\bibliography{bibliography}

\end{document}